\documentclass[10pt]{article}
\usepackage{geometry}                % See geometry.pdf to learn the layout options. There are lots.
\usepackage{graphicx}
\usepackage{subfigure}
\usepackage{setspace}
\usepackage{amssymb}
\usepackage{epstopdf}
\usepackage{amsmath}
\usepackage{color}
\usepackage{comment}
\usepackage{cite}

\newenvironment{en*}{\begin{eqnarray*}}{\end{eqnarray*}}

\begin{document}

\title{Enzyme oscillation can enhance the thermodynamic efficiency of cellular
metabolism: Consequence of anti-phase coupling between reaction flux and affinity}

\author{Yusuke Himeoka and Kunihiko Kaneko\\Department of Basic Science, University of Tokyo, Komaba, Meguro-ku, Tokyo 153-8902, Japan
}
\maketitle

\begin{abstract} Cells generally convert nutrient resources to useful products via energy
transduction. Accordingly, the thermodynamic efficiency of this conversion process is
one of the most essential characteristics of living organisms. However, although these processes occur under conditions of dynamic metabolism, most studies of cellular thermodynamic efficiency have
been restricted to examining steady states; thus, the relevance of dynamics to
this efficiency has not yet been elucidated. Here, we develop a simple model of metabolic
reactions with anabolism-catabolism coupling catalysed by enzymes. Through application of external oscillation in the enzyme abundances, the thermodynamic
efficiency of metabolism was
found to be improved. This result is in strong contrast with that observed in the
oscillatory input, in which the efficiency always decreased with
oscillation. This improvement was effectively achieved by separating the
anabolic and catabolic reactions, which tend to disequilibrate each other,
and taking advantage of the temporal oscillations so that each of the
antagonistic reactions could progress near equilibrium. In this case, anti-phase oscillation
between the reaction flux and chemical affinity through oscillation of
enzyme abundances is essential. This improvement was also confirmed in a model capable of generating
autonomous oscillations in enzyme abundances. Finally, the possible relevance of the
improvement in thermodynamic efficiency is discussed with respect to the potential for manipulation of metabolic oscillations in microorganisms.
\end{abstract}

%\begin{twocolumn}

\section{Introduction} 
~~~Cells uptake external nutrients from energy sources and transform them into all of the components required for growth, maintenance, and survival, such as the cell membrane and catalysts. The efficiency of these reaction processes, collectively referred to as "metabolism", is an important factor for the fitness of a cell. Given that intracellular reactions are catalysed by enzymes, the efficiency of such reactions also depends on the enzyme concentrations and their dynamics. Therefore, as cells regulate enzyme concentrations through protein expression dynamics, the potential relationship between metabolic efficiency and enzyme dynamics is an important issue warranting investigation \cite{maitra2015bacterial,kondo2011growth,himeoka2014entropy}. \\
~~~~ Although enzymes cannot alter the equilibrium condition itself \cite{bergethon2013physical}, they do change the speed of chemical reactions drastically. Indeed, enzymes generally facilitate chemical reactions in the order of $10^7$ to $10^{19}$ \cite{wolfenden2001depth}, and many reactions within a cell could
be almost completely halted by reducing the amount of the corresponding enzyme. In general, the relaxation process to equilibrium is controlled by the enzyme abundances, and thus so is the time scale for the metabolic reactions. \\
~~~~Furthermore, the abundance of each enzyme can change autonomously over time within a cell. Hence, together with the external flow that maintains the system out of equilibrium, the internal time scales are changed autonomously. In this sense, a cell is regarded as a machine with autonomous changes in time scale that functions in transforming nutrients into useful products via energy transduction \cite{himeoka2014entropy,kondo2011growth,segre2000compositional,kaneko2003recursiveness,stuart1993origins,jain2002large,hinshelwood1952136}. Although there are extensive studies on thermodynamic nature for a non-autonomous system that operates at a non-equilibrium condition with a finite velocity \cite{curzon1975efficiency,van2005thermodynamic,izumida2012efficiency,izumida2009onsager}, to date, the characteristic nature of such autonomous machinery has not been studied in the context of non-equilibrium chemical thermodynamics. Hence, investigations of the thermodynamic nature of dynamic metabolic processes are not only important to resolve basic questions in cellular biophysics but can also provide insight for non-equilibrium physics. \\
~~~Metabolic processes can be generally classified into anabolism and catabolism. The former is a synthesis process of biomolecules from nutrients, which typically involves consumption of energy by transforming ATP to ADP, whereas the latter involves decomposition of nutrients into smaller molecules, and consequently releases energy with a change from ADP to ATP. Through this metabolism, chemical resources are transformed into products along with the energy transduction between ATP and ADP. The thermodynamic efficiency of such energy transduction processes has been extensively studied \cite{westerhoff1982thermodynamics,russell1995energetics,rutgers1991control,westerhoff1983thermodynamic,farmer1976energetics,kayser2005metabolic,neijssel1996growth,vemuri2006overflow,von2006thermodynamics}. However, so far, these studies have primarily focused on the behaviours in the steady state, without consideration of the time-dependent dynamics in the reaction flux and concentrations of enzymes and substrates. \\
~~~~In contrast, there are a variety of time-dependent (non-steady) processes in cellular metabolism, such as those observed in the cell cycle \cite{morgan2007cell,murray1993cell}, cyclic AMP signaling \cite{martiel1987model}, circadian rhythms \cite{albert1997biochemical,hogenesch2011understanding} glycolytic oscillation \cite{duysens1957fluorescence,chance1964dpnh,ghosh1964oscillations,higgins1967theory,sel1968self,chance1967waveform,goldbeter1972dissipative,termonia1981oscillations}, and yeast metabolic cycle (YMC) \cite{tu2005logic,laxman2010behavior,tu2007cyclic,murray2007regulation,klevecz2004genomewide}. The enzyme concentrations change in time during these dynamic processes, which then alters the time scale of the reaction processes, as mentioned above. Thus, it is important to study the characteristics of such "dynamic" processes of cellular metabolism, and uncover the potential influence of the cyclic process on the thermodynamic efficiency of the metabolic processes. \\
~~~~Toward this end, we here introduce a simple model of coupled anabolic and catabolic reactions, each of which is catalysed by a corresponding enzyme and uses typical energy currency molecules (ATP and ADP). In particular, we demonstrate that temporal changes in enzyme concentrations are required to achieve higher thermodynamic efficiency under the condition in which available enzymes are limited.\\
~~~~In Section 2, the Nutrient, Waste, Substrate, Product (NWSP) model is described and implemented by applying periodic oscillations of the abundances of enzymes for anabolism and catabolism. This model showed improved thermodynamic efficiency compared to the steady-state condition. This result was then extended to the case with autonomous oscillations resulting from internal catalytic reaction dynamics, which confirmed the relevance of oscillation in enzyme abundances for thermodynamic efficiency. The observed increase in the efficiency due to the oscillation in enzyme abundances is in strong contrast with the general condition of a decrease in thermodynamic efficiency due to oscillatory inputs. In Section 3, we discuss the biological relevance of these results for the dynamic control of enzyme abundances.

\section{NWSP Model}
~~~Here, we introduce a simple model for a metabolic process consisting of anabolism and catabolism, by extending the model introduced by Westerhoff et al.\cite{westerhoff1982thermodynamics,russell1995energetics}. In the Westerhoff model, anabolism (process for the synthesis of biomolecules) and catabolism (process for the digestion of nutrients) are analysed using linear non-equilibrium thermodynamics, in which the deviation from chemical equilibrium is assumed to be small, and the steady chemical reaction flow and thermodynamic force (affinity) are proportional \cite{katzir1965nonequilibrium}. To incorporate oscillatory dynamics, we extended this model so that the chemical concentrations change according to the rate equation of the chemical reactions. Specifically, our model consists of four chemical species, i.e., Nutrient, Waste, Substrate, and Product, in addition to the energy currency molecules ATP and ADP, and involves two catalytic enzymes for anabolism and catabolism, $E_a$ and $E_c$, respectively. The catabolic reaction decomposes nutrients to waste with the aid of the catalyst $E_c$, simultaneously transforming ADP into ATP, whereas the anabolism reaction synthesizes a product from a substrate with the aid of $E_a$, by consuming energy with the change from ATP to ADP. As a consequence of the coupled reactions of catabolism and anabolism, the product is synthesised from a substrate by consuming a nutrient. If the enzyme concentrations are constant, there is a steady flow generated from nutrient and substrate to waste and product, depending on the concentrations of the chemical spoecies, as in the Westerhoff model. Here, we introduce a periodic change in the concentration of each enzyme, and study the effect of this temporal change in enzyme concentrations on the thermodynamic efficiency of this metabolism. Thus, to reflect all of these reaction dynamics, our model is given by
\begin{eqnarray}
\frac{{\rm d}[\rm{Nutrient}]}{{\rm d} t}&=&-J_c+\Bigl([\rm{Nutrient}]_{\rm{ext}}-[\rm{Nutrient}]\Bigr)\nonumber \\
\frac{{\rm d}[\rm{Waste}]}{{\rm d} t}&=&J_c+\Bigl([\rm{Waste}]_{\rm{ext}}-[\rm{Waste}]\Bigr)\nonumber \\
\frac{{\rm d}[\rm{Prod}]}{{\rm d} t}&=&-J_a+\Bigl([\rm{Prod}]_{\rm{ext}}-[\rm{Prod}]\Bigr)\nonumber \\
\frac{{\rm d}[\rm{Subs}]}{{\rm d} t}&=&J_a+\Bigl([\rm{Subs}]_{\rm{ext}}-[\rm{Subs}]\Bigr)\nonumber \\
\frac{{\rm d}[\rm{ATP}]}{{\rm d} t}&=&Ja+Jc \label {eq:ymc}\\
\frac{{\rm d}[\rm{ADP}]}{{\rm d} t}&=&-Ja-Jc\nonumber \\
J_c&=&[E_c](t)\Bigl([\rm{Nutrient}]\cdot[\rm{ADP}]-[\rm{Waste}]\cdot[\rm{ATP}]\Bigr)\nonumber \\
J_a&=&[E_a](t)\Bigl([\rm{Product}]\cdot[\rm{ADP}]-[\rm{Substrate}]\cdot[\rm{ATP}]\Bigr),\nonumber 
\end{eqnarray}	
where $[\cdot]$ represents the intracellular concentration of chemical species, and $[\cdot]_{\rm ext}$ represents the external concentration. If $E_c$ and  $E_a$ were constant, and the external concentrations of chemicals were close to be equilibrium values, the model system (\ref{eq:ymc}) would be reduced to that introduced by Westerhoff. Note that the total concentration of ATP and ADP is conserved according to (\ref{eq:ymc}). However, the enzyme concentration changes with time as
\begin{eqnarray*}
 [E_i]&=&\kappa(1+\sigma_i\lambda f(t))\\
\sigma_i&=&\begin{cases}
+1 & (i=c)\\
-1 & (i=a),
\end{cases}
\end{eqnarray*}
where $f(t)$ is a given periodic function with period $T$ that satisfies,
\begin{itemize}
 \item $\max[f(t)]=1,\min[f(t)]=-1$
 \item$\int_{0}^Tf(t){\rm d} t =0$.\end{itemize}
With the first condition, $\lambda$ gives the amplitude of the oscillations in the enzyme concentration (relative to the steady-state value). The second condition is imposed so that the average concentration is not altered by the periodic change, which facilitates comparison between the steady and oscillatory cases. The average concentration of enzymes is given by the parameter $\kappa$, which controls the time scale of the chemical reactions. The sign parameter $\sigma_i$ is introduced to represent the difference in the phase of enzyme oscillations between catabolism and anabolism. Here, we mainly consider the case in which catabolism and anabolism proceed in an anti-phase manner, and the case with an in-phase will be briefly discussed later.
\begin{figure}[htbp]
\begin{center}
  \includegraphics[width = 100 mm,angle=0]{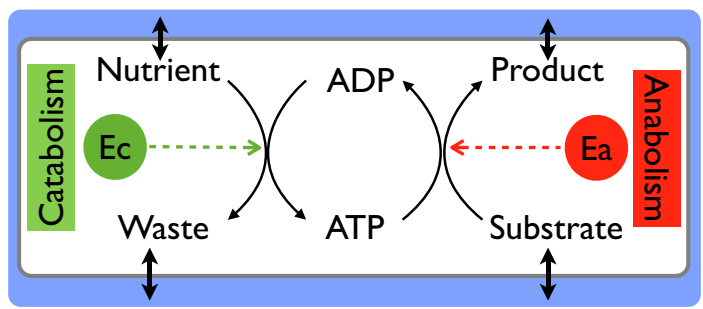}
  \caption{Schematic representation of the NWSP model. Nutrient, Waste, Product, and Substrate molecules diffuse in and out of the cell through the external environment. Enzyme concentrations $[E_c]$ and $[E_a]$ are periodically changed externally in order to study the influence of oscillations on the thermodynamic efficiency.}
  \end{center}
\label{fig:YMC_toymodel}

\end{figure}

\subsection{Remarks on oscillatory input}
~~~~Before presenting the results for the enzyme oscillation, we provide brief remarks on the general consequence of oscillations in substrate concentration.\\
~~~~In general, cellular metabolism can be described as a transduction system of chemical energy. Therefore, it can be expected that oscillatory reaction dynamics would increase the dissipation in energy transduction compared with the steady-state case. For example, let us consider the simplest reversible chemical reaction $X\leftrightarrow Y$, with $X$ as the substrate and $Y$ as the product. Then, it can be clearly proven with linear non-equilibrium dynamics that dissipation in the chemical reaction represents a minimum for the steady state (for details, see the Supplement). In general, such oscillations do not change the time scale for the equilibration, and never improve the thermodynamic efficiency.\\
~~~~In contrast, the enzymes themselves can change the time scale for the reaction, which facilitates the equilibration process. Hence, with appropriate oscillatory dynamics in enzyme concentrations, higher thermodynamic efficiency may be achieved, which we will explore in this section. \\
\subsection{Characteristic dynamics}
\begin{figure}[htbp]
  \includegraphics[width = 160 mm,angle=0]{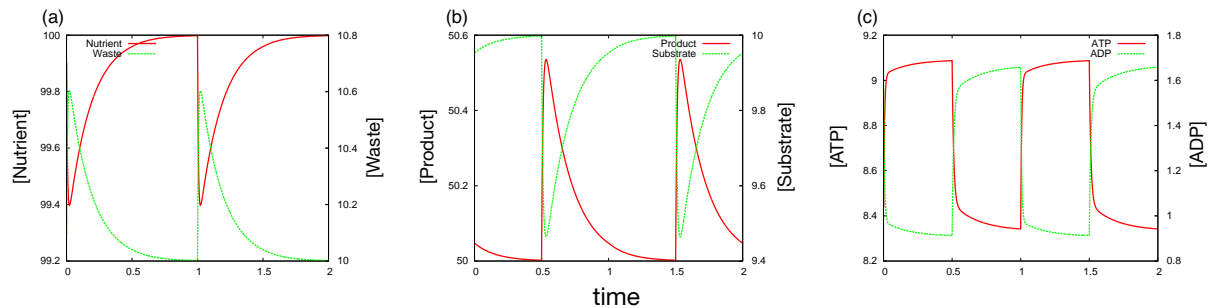}
  \caption{An example of the time series of chemical concentrations in the NWSP model (\ref{eq:ymc}) plotted over two periods. The periodic function $f(t)$ is chosen to be a step function with period $1$. We set ${\rm[Nutrient]_{\rm ext}}=100.0,{\rm[Waste]_{\rm ext}}=10.0,{\rm[Product]_{\rm ext}}=50.0,{\rm[Substrate]_{\rm ext}}=10.0,{\rm [ATP,ADP]_{\rm total}}=10.0$, and $\kappa=0.1$.}
  \label{fig:time_series}
\end{figure}

~~~~Figure {\ref{fig:time_series}} shows an example of the time series of the concentrations of nutrients, substrates, waste, product, ATP, and ADP, where the periodic function $f(t)$ is chosen as successive switches by a step function with period $1$. Specifically, for $t\in [0.0,0.5)$, $[E_c]=\kappa$ and $[E_a]=0.0$, and for $t\in [0.5,1.0)$, $[E_c]=0.0$ and $[E_a]=\kappa$, where $[E_c]$ and $[E_a]$ change periodically. In the example given in Figure 2, the enzyme is switched four times at $t=0.0,0.5,1.0$, and $1.5$. Changes in the concentrations of ATP and ADP reflect this enzyme switching, whereas the concentrations of the nutrient and waste are not changed substantially at $t\approx 0.5$ (and $t\approx 1.5$). This is why the chemical reaction Nutrient $+$ ADP $\leftrightarrow$ Waste $+$ ATP is almost relaxed to equilibrium at that time point, and the affinity of the reaction, i.e. the flux, is approximately zero. On the other hand, the product and substrate do not change at $t\approx 0.0$ (and $t\approx 1.0$) for the same reason.\\
~~~~Here, we introduce the average thermodynamic efficiency of metabolism to study the relevance of oscillatory metabolism for energy transduction. This model of thermodynamic efficiency was originally introduced by Westerhoff et al. \cite{westerhoff1982thermodynamics,russell1995energetics} as a ratio of the output of Gibbs free energy to the input of Gibbs free energy, given by $\eta=-J_aA_a/J_c A_c$, where $J_i$ and $A_i$ are the chemical reaction flux and affinity (difference in the Gibbs free energy between the substrate and product of the reaction), respectively \cite{kedem1965degree}. However, this model of the thermodynamic efficiency of metabolism was originally considered in the steady state without a temporal change in chemical concentrations. Therefore, to deal with the case of a time-dependent chemical reaction system, we here extend the definition of the thermodynamic efficiency as follows:
\begin{eqnarray}
\eta(\lambda,\kappa)&=&-\frac{\int_0^TJ_a(t)\cdot A_a(t){\rm d} t}{\int_0^TJ_c(t)\cdot A_c(t) {\rm d} t} \nonumber \\
A_c&=&G_{\rm nutrient}-G_{\rm waste} \label{eq:def_ymc} \\
A_a&=&G_{\rm product}-G_{\rm subst}, \nonumber
\end{eqnarray}
 where $G_i$ is the chemical potential (sum of the standard chemical potential and activity due to a difference in concentration from the standard) of chemical species $i$, $\kappa$ is the rate constant of the chemical reaction, and $\lambda$ is the amplitude of enzyme oscillations relative to the steady state. By setting $\lambda=0$ (without oscillation), this definition (\ref{eq:def_ymc}) is of the same form used for the steady-state metabolism model \cite{westerhoff1982thermodynamics}; thus, it is a natural extension to incorporate the oscillatory case.\\
~~~~Rigorously speaking, this definition of thermodynamic efficiency is justified only near equilibrium. Nevertheless, it is expected that $\eta$ provides at least an approximately good measure for the efficiency of chemical energy transduction in consideration of oscillatory reaction dynamics.
\subsection{Influence of oscillation on the thermodynamic efficiency}
\begin{figure}[htbp]
\begin{center}
  \includegraphics[width = 100 mm,angle=0]{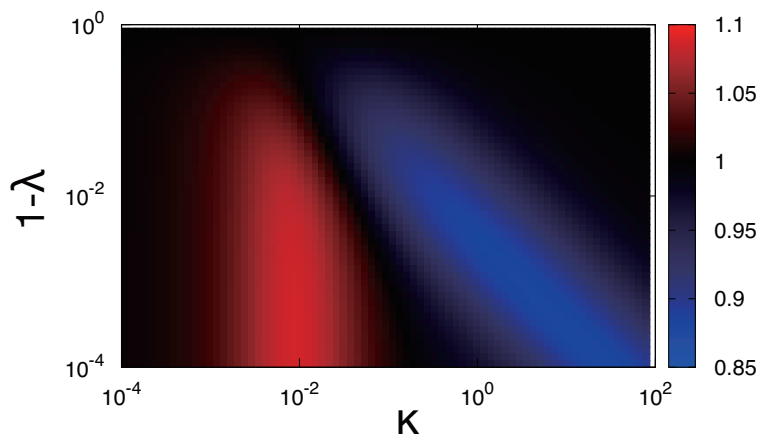}
    \caption{Heat map of the average thermodynamic efficiency with oscillations in enzyme concentrations relative to that in the steady state. The horizontal and vertical axes are $\kappa$ and $1.0-\lambda$, respectively. The colour bar represents $\eta(\lambda,\kappa)/\eta(\lambda=0,\kappa)$. The thermodynamic efficiency $\eta$ is improved for the small $\kappa$ region. The step periodic function, with period $T=200.0$, was adopted to model the change in enzyme concentration. Parameters were set as follows: ${\rm [Nutrient]_{\rm ext}}=10.0$, ${\rm [Waste]_{\rm ext}}=0.1$, ${\rm [Prod]_{\rm ext}}=1.0$, ${\rm [Subs]_{\rm ext}}=0.05$,and ${\rm [ATP]+[ADP]}=100.0$.}
  \label{fig:heatmap}
  \end{center}
\end{figure}

\begin{figure}[htbp]
\begin{center}
\includegraphics[width = 100 mm, angle = 0]{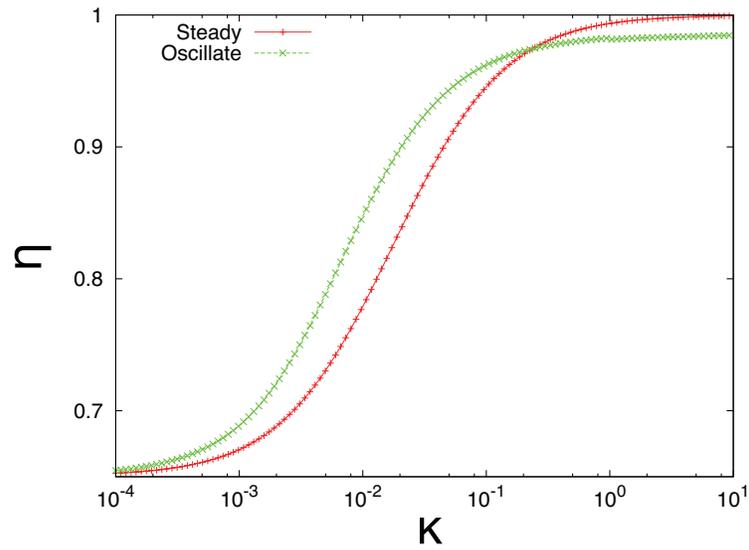}
  \caption{The average thermodynamic efficiency $\eta(\lambda,\kappa)$ for the steady ($\lambda=0$) and oscillatory ($\lambda=1$) cases plotted against the rate constant of the chemical reaction $\kappa$. The average thermodynamic efficiency for the oscillatory case is higher for the small $\kappa$ region. The choice of function and parameters is identical with that described in Figure\ref{fig:heatmap}.}
    \label{fig:kappa}
  \end{center}
\end{figure}

~~~~In this subsection, we demonstrate the dependence of $\eta$ on the amplitude $\lambda$ of the oscillation of enzyme concentrations and their average $\kappa$ values. The efficiency is plotted as a function of the amplitude $\lambda$ and rate constant $\kappa$ in Figure \ref{fig:heatmap}. 
Here, the steady state is given by $\lambda=0$, while $\lambda=1$ corresponds to the switch between two separated states, in which one of the enzymes ($E_a$ or $E_c$) vanishes so that the corresponding reaction is halted. From the numerical results in Figure \ref{fig:heatmap}, we find that the incorporation of oscillatory metabolism improves the average efficiency when $\kappa$ is small. However, note that the average flux, $<J>(\lambda,\kappa)=-T^{-1}\int_0^TJ_a(t){\rm d} t$, is not improved (see Figure S2).\\
~~~~In Figure \ref{fig:kappa}, the efficiency is compared between the steady ($\lambda=0$) and full oscillatory ($\lambda=1$) cases. This comparison further confirmed that the efficiency $\eta(\lambda,\kappa)$ is improved with oscillations in the small $\kappa$ region. Note that $\eta(\lambda,\kappa)$ is not a monotonic function of $\lambda$ for a certain range of $\kappa$ values. The crossover point of the amplitude at which the efficiency for the oscillation case exceeds that for the steady case depends on $\kappa$, whereas the oscillation case is always advantageous as long as $\kappa$ is small. This advantage of dynamic (chemical) energy transduction for thermodynamic efficiency is in strong contrast with the energy conversion observed in the dynamic input change, in which the efficiency is always decreased by the oscillation, as discussed above.
\\
~~~In the following, we will focus on how this observed improvement in efficiency is actually achieved. Firstly, loss in efficiency is caused by the dissipation of Gibbs free energy in each chemical reaction. Then, the amount of dissipation is reduced when each reaction progresses close to its equilibrium. If $\kappa$, the time scale for the catalytic reaction, is sufficiently larger than that of material flux (which is set to unity), the system is near equilibrium and the loss is small; indeed, this loss is smallest at the steady state. However, when $\kappa$ is small, the "chemical coupling`` between the two reactions (anabolism and catabolism) hinders the approach to equilibrium. In the present case, the two reactions are coupled with the energy currency ATP and ADP. Equilibration of catabolism results in an excess of ATP compared to ADP, which results in a far-from-equilibrium condition for the anabolism reaction (Of course, whether equilibration of one elemental reaction dis-equilibrates the other will depend on how the two reactions are coupled energetically; in particular, this will depend on the values of standard chemical potentials. However, as long as $\int J_c {\rm d} t>0,\int J_a{\rm d} t<0$ and $G_{\rm Substrate}<G_{\rm Product}$, synthesis of the product (P) relies on the free energy of ATP synthesised from the nutrient, and this coupling form is generally true for coupling between catabolic and anabolic reactions. Thus, increasing the reaction time scale of catabolism produces more ATP, which then dis-equilibrates the anabolic reaction.).\\
~~~Thus, the elemental reaction processes dis-equilibrate each other in the present coupled reaction system. In the steady state, these two reactions progress in a moderately dis-equilibrated condition when $\kappa$ is small. In contrast, the oscillatory reaction can separate the process into two time regimes. The first regime is the situation for $E_c$, in which the catabolism reaction progresses near equilibrium and the anabolism reaction occurs far from equilibrium and is almost halted as $E_a\sim 0$. On the other hand, in the time regime with $E_a>0 $ and $E_c\sim 0$, anabolism progresses near equilibrium, and catabolism is almost halted. Thus, most of the reaction events for the two temporal regimes occur near their equilibria, and thus the dissipation is suppressed in these two regimes. On the other hand, during the switch time between $E_c\gg E_a$ and $E_a\gg E_c$, the dissipation would be increased. Thus, the total efficiency depends on the difference between the gain in the suppression from one of the reactions and the loss caused by the oscillation. This characteristic is not observed when the enzymes oscillate in an in-phase manner. In the in-phase oscillation, the enzymes facilitate the catabolic and anabolic reactions equally, and each reaction hinders the approach to the other reaction's equilibrium, as observed in the steady state.\\
~~~~To quantify the gain and loss of this oscillatory chemical reaction system, we calculated the energy gain $G_{\rm gain}$ and loss $G_{\rm loss}$ using the following equation:  
\begin{eqnarray}
G_{\rm gain}&=&\sum_{i=c,a}\int_{0}^T\theta(A^{\rm st}_i-A^{\rm os}_i(t))J_i^{\rm os}(t)(A^{\rm st}_i-A^{\rm os}_i(t)){\rm d} t\nonumber \\
G_{\rm loss}&=&\sum_{i=c,a}\int_{0}^T\theta(A^{\rm os}_i(t)-A^{\rm st}_i)J_i^{\rm os}(t)(A^{\rm os}_i(t)-A^{\rm st}_i){\rm d} t\label{eq:loss_and_gain}\\
\theta(x)&=&\begin{cases}
1 & (x>0)\nonumber\\
0 & (x<0)\nonumber,
\end{cases}
\end{eqnarray}
where $A_i^j$ is the affinity (Gibbs free energy difference) of the reaction, and $J_i^j$ is the corresponding flux, where $i=c$ for catabolism and $i=a$ for anabolism; and $j={\rm st}$ for $\lambda=0$ (the steady case) and $j={\rm os}$ for $\lambda=1$ (full oscillatory case). Thus, $G_{\rm gain}$ ($G_{\rm loss}$) represents the amount of reduced (excess) Gibbs free energy compared with that of the steady-state value. The value of $\kappa$ at which the gain exceeds the loss is approximately equal to the value where $\eta(\lambda=1,\kappa)$ is larger than $\eta(\lambda=0,\kappa)$, as shown in Figure \ref{fig:kappa}. In summary, by progressing the chemical reaction closer to the equilibrium and suppressing the reaction further from equilibrium, the efficiency is increased compared to that of the steady case.  \\
~~~~This improvement in efficiency can also be interpreted from a different perspective. Figure \ref{fig:anti_phase_NWSP} shows the time series of the flux and affinity of the catabolic reaction. When the flux is not close to zero, the affinity is close to zero, and thus the reaction progress near equilibrium. By contrast, when the affinity is not close to zero, the flux is close to zero, except at the time point of enzyme switching, and this far-from-equilibrium reaction occurs only at a very low rate. Thus, thermodynamic loss is suppressed throughout the process. In Figure \ref{fig:anti_phase_NWSP}, we illustrate this negative correlation between flux and affinity for catabolism, which was also confirmed for anabolism.\\
~~~~It is important to note that this characteristic cannot be realized under the condition of linear thermodynamics without a change in the enzyme concentrations, in which thermodynamic flux increases with the affinity. Thus, the flux can be controlled only by changing the affinity of the chemical reaction, and the two are always positively correlated. In contrast, in the present case, by changing the concentration of the catalytic enzyme, the time scale of the chemical reaction is changed. Accordingly, the flux can be increased even for a low-affinity state, and can be suppressed for a high-affinity state. \\
\begin{figure}[htbp]
\begin{center}
\includegraphics[width = 100 mm, angle = 0]{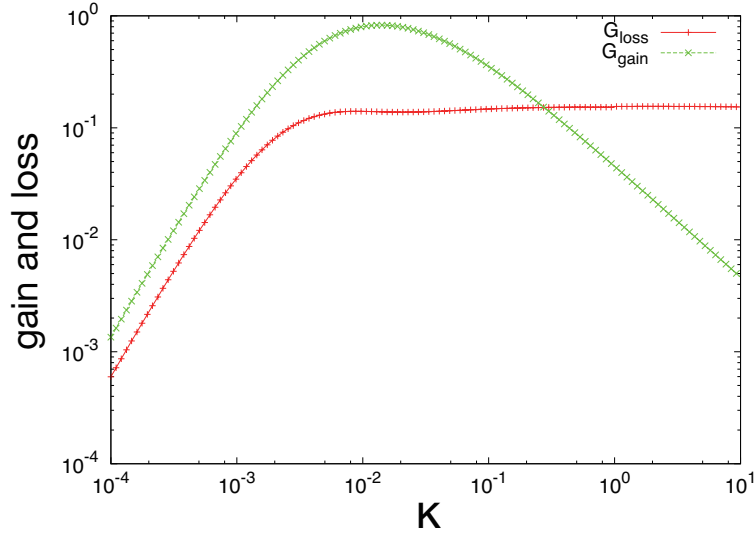}
  \caption{The thermodynamic loss and gain with oscillatory metabolism calculated using (\ref{eq:loss_and_gain}). At the value of $\kappa$ for which the loss and gain are balanced, $\eta(\lambda=0,\kappa)\approx\eta(\lambda=1,\kappa)$ holds as in Figure \ref{fig:kappa}. The choice of the periodic function and parameter values is identical with that of Figure \ref{fig:heatmap}.}
    \label{fig:benefit} 
  \end{center}
\end{figure}
\begin{figure}[htbp]
\begin{center}
\includegraphics[width = 100 mm, angle = 0]{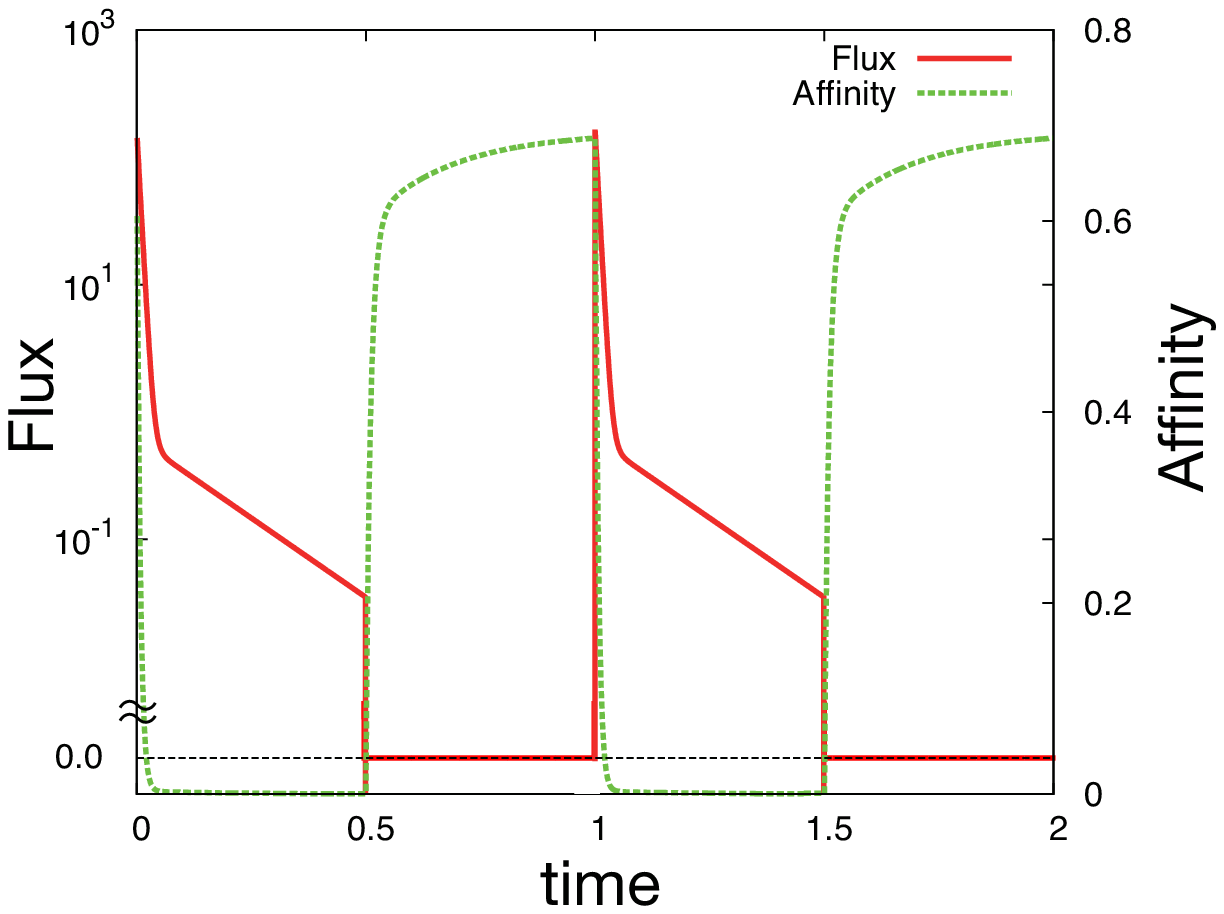}
  \caption{Time series of the flux (red) and affinity (green) under enzyme switching plotted over two time periods. A step periodic function with period $T=1$ was adopted for $f(t)$.The flux and affinity ocillate with an anti-phase.}    
    \label{fig:anti_phase_NWSP}
  \end{center}
\end{figure}
\section{Extension to an autonomous chemical reaction model}
\begin{figure}[htbp]
\begin{center}
\includegraphics[width = 60 mm, angle = 0]{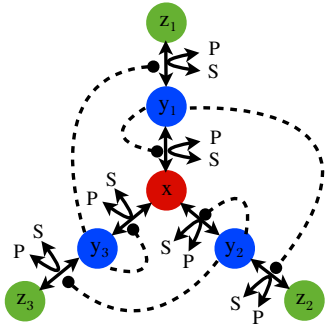}
  \caption{Schematic representation of a simple chemical reaction model exhibiting autonomous oscillation. The chemical reaction $x\leftrightarrow y_i$ is catalysed by $y_i$, and $y_i \leftrightarrow z_i$ is catalysed by $y_{i-1}$ species. In each reaction, the product (P) is synthesised from substrates (S) that are externally supplied. An anabolic reaction $substrate\leftrightarrow product$ is coupled to each reaction.}
    \label{fig:repressilator}
  \end{center}
\end{figure}

\begin{figure}[htbp]
\begin{center}
\includegraphics[width = 170 mm, angle = 0]{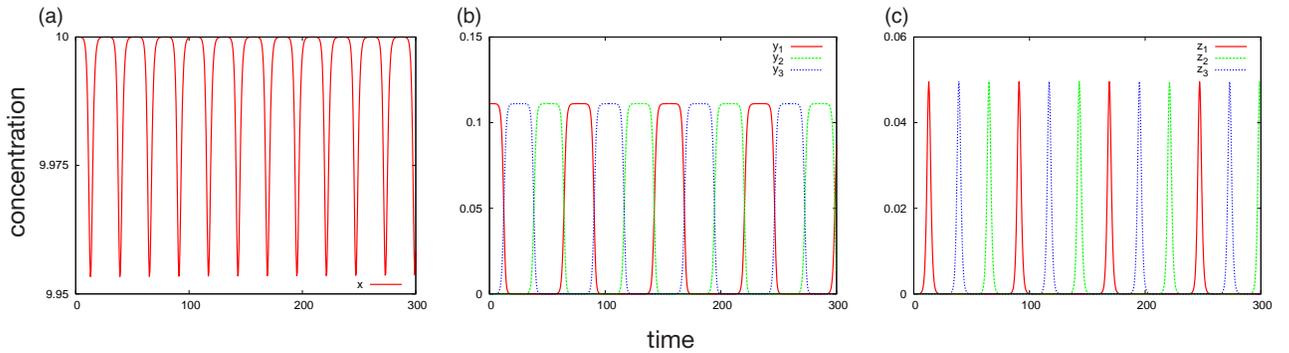}
    \caption{Time series of concentrations of species (a) $x$, (b) $y_i$s, and (c) $z_i$s. Parameters were chosen as follows: $\mu_x=1.0,\mu_y=2.0,\mu_z=0.0,\mu_{\rm product}-\mu_{\rm substrate}=0.5,\kappa_1=0.1,\kappa_2=20.0,\beta=3.0,D=\phi=1.0$, $[X]=10.0$, and $[Z]=0.0$.}
    \label{fig:timeseries_replessilator}
  \end{center}
\end{figure}

~~~~So far we have demonstrated that the thermodynamic efficiency of metabolism (coupling between anabolism and catabolism) can be improved by incorporation of an oscillatory chemical reaction. However, the model introduced in the last section cannot generate the chemical oscillation autonomously; thus, this model does not account for the possible energetic cost from generating the chemical oscillation. Since the chemical oscillation is generated as a dissipative structure at the far-from-equilibrium condition, generating the oscillation itself requires a certain amount of excess dissipation \cite{cao2015free}. Therefore, it is important to investigate whether or not a chemical reaction system capable of generating autonomous oscillation can improve thermodynamic efficiency. To address this question, we developed a three-catalyst model in which chemical oscillation appears under a given flow from resource chemicals to product.\\
~~~~Our model consists of nine chemical species, $x, y_i, z_i, (i=1,2,3)$, substrate, and product. Chemical reactions were constructed by coupling between catabolic (energy-generating) and anabolic (energy-consuming) reactions, as shown in the schematic in Figure \ref{fig:repressilator}. There are two types of catalytic reactions, $x+{substrate}\leftrightarrow y_i+{ product}$ and $y_i+{substrate}\leftrightarrow z_i+{product}$, and each reaction is catalysed by $y_i$. The system takes up $x$ and the substrate from the external environment, and $x$ is converted into $z$ via the above two reactions, resulting in the simultaneous synthesis of two products. The produced $z$ molecules and products are transported into the environment. In the model, the reaction to covert $x$ into $z$ via $y$ is assumed to be a catabolic reaction, and the chemical reaction between the substrate and the product is regarded as anabolism. Therefore, the reaction ${ substrate} \rightarrow { product}$ is driven by the coupled chemical reactions $x\rightarrow y_i$ and $y_i \rightarrow z_i$.\\         
~~~~In developing this model, we referred to the example of a repressilator \cite{elowitz2000synthetic}, in which the expression of three proteins is mutually inhibited to generate the oscillation. However, given the present focus of a reaction system involving a catalytic reaction, there is no direct inhibition. Instead, each of the three components $y_i$  mutually catalyse the decomposition of $y_{i+1}$ into $z_{i+1}$ ($y_{i+1}+{substrate }\leftrightarrow z_{i+1}+{ product}$). Hence, the $y_i$s suppress the abundances of each other, which introduces effective mutual inhibition, as in a repressilator. \\  
~~~~Here, for the sake of simplicity, we assume that the consumed substrate or synthesised product is quickly supplied by or transported to the external environment, respectively; thus, their internal concentrations are kept constant. Therefore, there are only $7$ variables for $[x]$, $[y_i]$, and $[z_i]$. In addition, by scaling the rate constants according to the concentration of the substrate, the coupling with the anabolic reaction is given by a single parameter $\rho$. Finally, by assuming the symmetrical case in which the parameters for each species $i$ are homogeneous over the index $i$, our chemical reaction system is given by
\begin{eqnarray}
 \frac{{\rm d} [x]}{{\rm d} t}&=&-\sum_{i=1}^3J_{1,i} + D([X]-[x])\nonumber \\
 \frac{{\rm d} [y_i]}{{\rm d} t}&=&J_{1,i}-J_{2,i}\nonumber \\
 \frac{{\rm d} [z_i]}{{\rm d} t}&=&J_{2,i} +\phi([Z]-[z_i])\label{eq:repressilator}\\
 J_{1,i}&=&\kappa_1[y_i]([x]-l_1 \rho [y_i])\nonumber \\
 J_{2,i}&=&\kappa_2[y_{\sigma(i)}]([y_i]-l_2 \rho [z_i]),\nonumber
\end{eqnarray}
where $[\cdot]$ represents the concentration of the corresponding chemical species, $\kappa_1$ and $\kappa_2$ are the rate constants of each chemical reaction $x+substrate \leftrightarrow y_i +product$ and $y_i + substrate + z_i+product$, respectively, $D$ and $\phi$ are the rate of material exchange of $x$ and $z$ with the external environment, respectively, and $l_1$, $l_2$, and $\rho$ are the Boltzmann factors of each chemical reaction (i.e. $l_1=\exp(-\beta(\mu_x-\mu_y))$, $l_2=\exp(-\beta(\mu_y-\mu_z))$, and $\rho=[{\rm product}][{\rm substrate}]^{-1}\exp(-\beta(\mu_{\rm substrate}-\mu_{\rm product}))$, where $\mu_i$ is the standard chemical potential of chemical species $i$). $[X]$ and $[Z]$ represent the environmental concentrations of chemical species $x$ and $z$, respectively, where $\sigma$ denotes the cyclic permutation on $i$ ($\sigma(1)=3,\sigma(2)=1,\sigma(3)=2$). \\
~~~~We numerically computed the dynamics and attractor of the model by varying the parameter values, and found that the system exhibits Hopf bifurcation from a fixed point to limit-cycle oscillation with the increase in the parameter $\kappa_2$. An example of the time series of the concentrations is shown in Figure \ref{fig:timeseries_replessilator}.\\
\subsection{Relationship between oscillation and dissipation}
\begin{figure}[htbp]
\begin{center}
   \includegraphics[width = 150 mm, angle = 0]{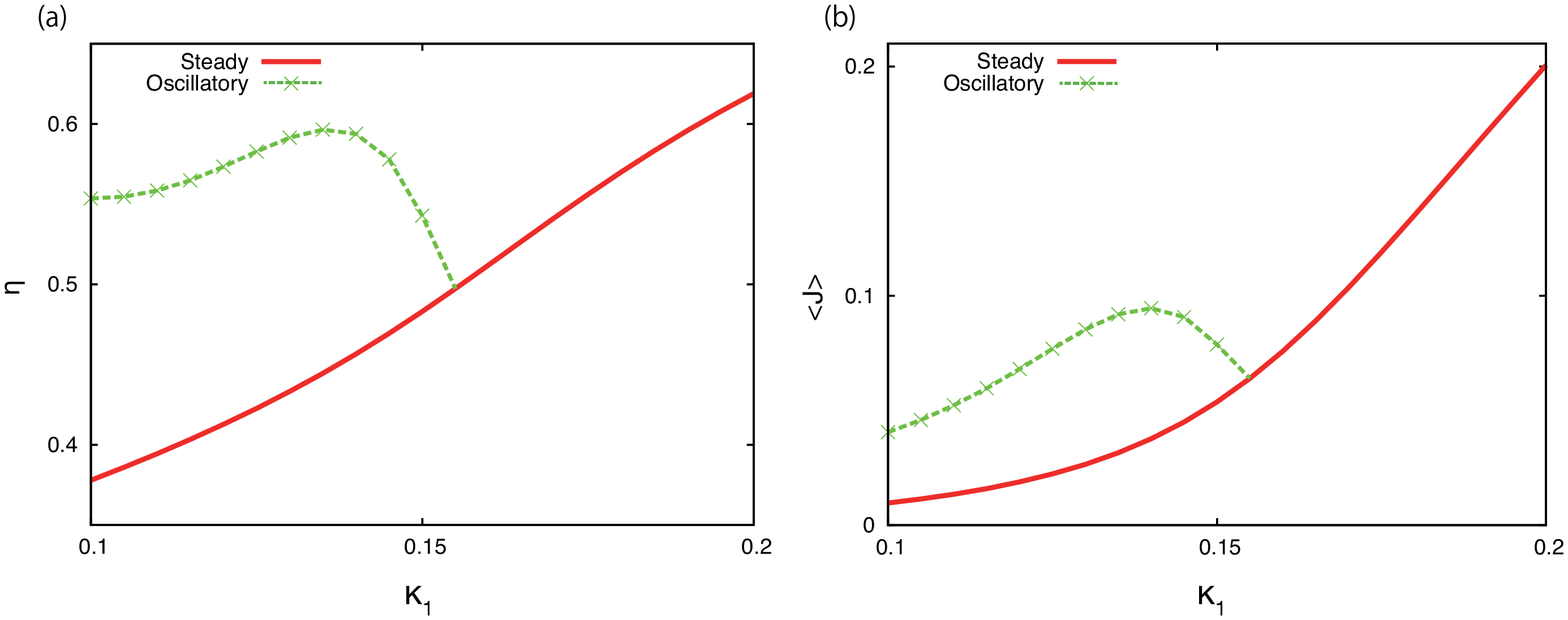}
    \caption{(a) The thermodynamic efficiency and (b) average flux of the model, plotted as a function of $\kappa_1$, for the oscillation (green) and steady (red) cases. The thermodynamic efficiency is maintained at a higher value by the oscillation for small $\kappa_1$ values. In addition, the oscillation also improved the flux. Hopf bifurcation occurs at $\kappa_1 \approx 0.155$, and the oscillatory solution (limit cycle) disappears for $\kappa_1>0.155$. The parameters were chosen to be $\mu_x=1.0,\mu_y=2.0,\mu_z=0.0,\mu_{\rm product}-\mu_{\rm substrate}=1.5,\kappa_2=250.0,\beta=1.0,D=\phi=1.0$, $[X]=10.0$, and $[Z]=0.0$.}
    \label{fig:compare}
  \end{center}
\end{figure}

\begin{figure}[htbp]
\begin{center}
\includegraphics[width = 150 mm, angle = 0]{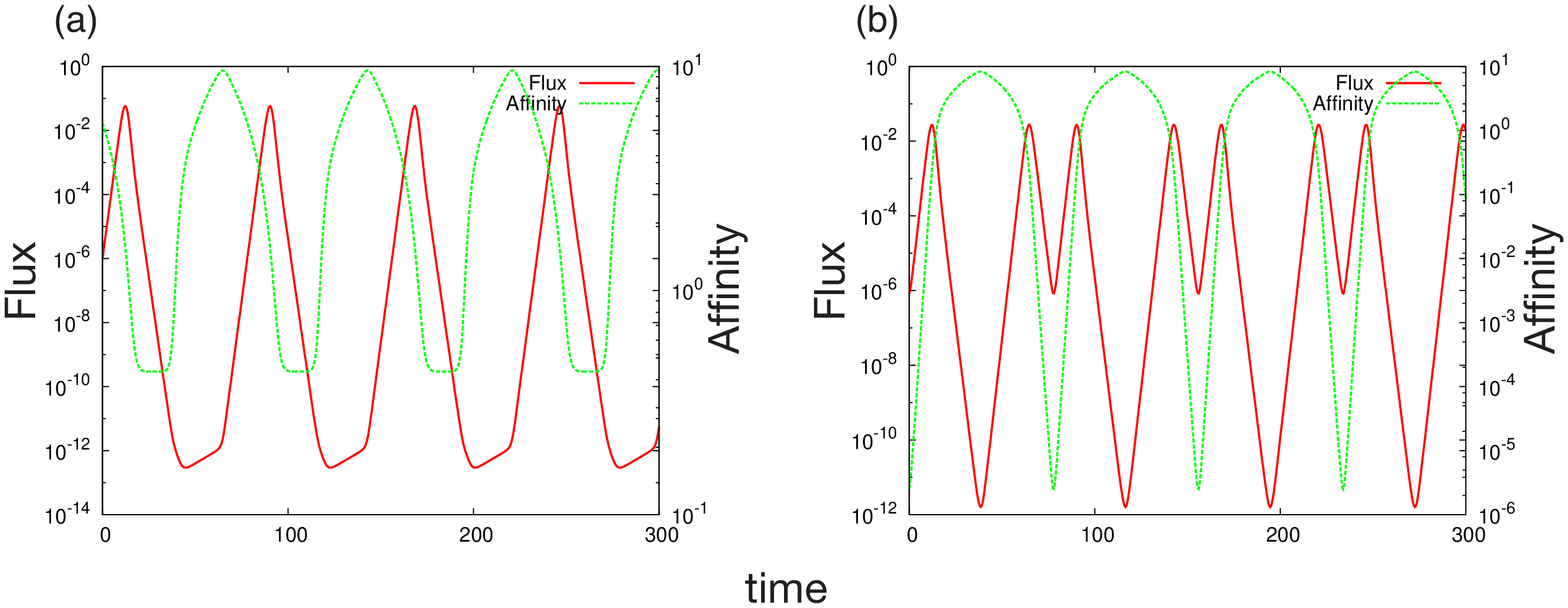}
     \caption{Time series of the flux (red) and affinity (green) in the reaction $x+substrate\leftrightarrow y_1+product$ (a), and $y_1 +substrate \leftrightarrow z_1+product$ (b). The flux and affinity of both reactions, $x+substrate \leftrightarrow y_1+product$ and $y_1+substrate\leftrightarrow z_1+product$, oscillate out of phase as in the NWSP model. The time series for index $i=1$ is shown here, and those for indices $i=2$ and $i=3$ are the same (except for the difference in the oscillation phase). The parameter values were chosen as described in Figure \ref{fig:timeseries_replessilator}.}
    \label{fig:expand}
  \end{center}
\end{figure}

~~~~We now introduce the average thermodynamic efficiency of the energy transduction, as described in Section 2. In our model (\ref{eq:repressilator}), the reactions to convert $x$ into $z$ via $y$ are regarded as catabolic, and the reactions between the substrate and product are considered to be anabolism. Thus, the average thermodynamic efficiency can be defined as,
$$\eta=-\frac{\sum_{i,j}\int_0^TJ_{i,j}(t)A_{i,j}^a{\rm d} t}{\sum_{i,j}\int_0^TJ_{i,j}(t)A_{i,j}^c(t){\rm d}
 t},$$
where $A_{i,j}^c\ (A_{i,j}^a)$ indicates the affinity of the catabolic (anabolic) part of the chemical reaction $(i,j)$, and the time for the average $T$ is chosen to be sufficiently long. Specifically, the parameters are given by $A_{1,j}^c=\ln(x/(l_1\cdot y_j))/\beta,A_{2,j}^c=\ln(y_j/(l_2\cdot z_j))/\beta$, and $A_{i,j}^a=-\ln(\rho)/\beta$. The definition of the average thermodynamic efficiency implies the ratio of output to input energy, and $\eta$ satisfies the inequality $0\leq \eta \leq 1$ with appropriate conditions: $J_{i,j}(t)$ values are always positive, and $\rho\geq1$.\\
~~~~Figure \ref{fig:compare} shows the average flux $<J>$ and $\eta$. Both the average flux and the thermodynamic efficiency were improved (or maintained at a high level) by incorporation of the oscillation. The time courses of the flux and affinity for each reaction are shown in Figure  \ref{fig:expand}. The flux and affinity of each chemical reaction oscillates out of phase, roughly in anti-phase. Since the thermodynamic dissipation is given by the product between the flux and affinity, it can be reduced by this anti-phase oscillation between the two, as described in Section 2.\\
~~~~To summarise, we introduced a simple chemical reaction system exhibiting autonomous sustainable oscillations, and confirmed that the oscillations increased the average thermodynamic efficiency, which was attributed to the anti-phase oscillation between the flux and affinity.\\   
\section{Summary and Discussion}
~~~~We here report the first evaluation of the relevance of oscillatory chemical reactions to thermodynamic efficiency by introducing the NWSP model. In the model, chemical energy is stored by transforming ADP into ATP via a catabolic reaction that converts an imported nutrient into waste, while an anabolic reaction consumes the stored energy to synthesise a product from a substrate, by transforming ATP to ADP. Such a coupling structure between anabolism and catabolism forms the basis of cellular metabolism. Thus, the NWSP model is regarded as a simplified, coarse-grained model for cellular metabolism. Each reaction is catalysed by a corresponding enzyme. Moreover, by introducing oscillations in the abundances of each enzyme, the thermodynamic efficiency of energy conversion was computed and compared between the steady and oscillatory cases.\\
~~~~Under the condition of enzyme limitation, in which the (average) enzyme abundances are insufficient and the chemical reactions can only progress slowly, we found that the oscillation in  enzyme abundances improved the thermodynamic efficiency. This improvement stems from the antagonistic nature between anabolism and catabolism. In general, a catabolic reaction generates the energy currency by breaking a large macromolecule into smaller pieces, whereas anabolism consumes energy to drive the synthesis of a macromolecule. Therefore, these two reaction types tend to dis-equilibrate each other, and thus they will progress far from equilibrium if they progress concurrently under a limited rate, which will consequently increase thermodynamic loss (entropy production). In contrast, if  sequential switching between anabolism and catabolism is achieved by oscillations in enzyme abundances, the two reactions become decoupled, and the loss is suppressed. Indeed, we confirmed the improvement in thermodynamic efficiency by imposing anti-phased oscillations in the abundances of enzymes for anabolism and catabolism. \\
~~~~However, the generation of such oscillation can bring about thermodynamic loss by itself. To account for this effect, we also considered a reaction model that can generate autonomous chemical oscillations, by amending the energy-transduction with three enzymes that mutually catalyse their degradation reactions, so that autonomous, repressilator-type, oscillation is possible. The model converts chemical species $x$ into $z$ via the enzymes, along with synthesis of a product from the substrate. By comparing the thermodynamic efficiency between cases, we confirmed again that the thermodynamic efficiency is improved by the emergence of chemical oscillation.\\
~~~~The common mechanism for the improvement in efficiency observed in our two models is the anti-phase oscillation between the chemical reaction flux and affinity. The anti-phase oscillation implies low flux for a high-affinity state and high flux for a low-affinity state. This situation leads to the increase in the thermodynamic efficiency, as the thermodynamic loss owing to entropy production is given by the product between the flux and affinity.\\
~~~~In the present study, we focused only on a system with simplified, coarse-grained metabolism and elementary chemical reactions. However, improvement in the thermodynamic efficiency by such anti-phase oscillation between flux and affinity is expected to be a universal phenomenon for all biochemical processes involving catalytic reactions. In the case of an ordinary chemical reaction without catalytic enzymes, chemical reaction flux can only be controlled by the affinity, and the two are positively correlated (or proportional in the case of linear thermodynamics) \cite{kacser1993universal,kacser1973control,katzir1965nonequilibrium}. However, for an enzymatic reaction, the flux is controlled not only by the affinity but also by the abundances of the corresponding enzymes. The enzymes do not directly alter the equilibrium condition but do facilitate the tendency toward equilibration. Hence, an increase in enzyme supply reduces dissipation for a single chemical reaction event \cite{himeoka2014entropy}. Therefore, as long as the enzyme abundance is increased with the flux, the affinity and flux will be negatively correlated if the enzyme controls the chemical reaction.\\
~~~~In general, biochemical
reactions in microbial experiments are often facilitated and halted by simply controlling the amount of
enzymes.\cite{scott2010interdependence,molenaar2009shifts,stouthamer1973utilization,iglesias2010control,heinrich1974linear,kacser1993universal,kacser1973control}.
Therefore, the results of the present model suggest that thermodynamic efficiency in metabolism could be
improved by dynamically changing the enzyme concentrations in a microbial system.
For example, oscillations in gene expression have been reported in yeast grown under
nutrient limitation, which is generally referred to as the yeast metabolic cycle
(YMC)\cite{tu2005logic,laxman2010behavior,tu2007cyclic,murray2007regulation}. Roughly speaking, expression levels of anabolic and
catabolic proteins are temporally separated in the YMC, in a similar manner
to the condition of our NWSP model under an oscillatory switch of the two enzymes. Here, as
the nutrients are limited, the rate of the catalytic reaction is lowered, which implies a smaller $\kappa$ value (the rate constant of chemical reactions) in our NWSP
model. Hence, improvement of the thermodynamic efficiency by
oscillation for the small $\kappa$ region observed here may provide an explanation
for the experimental observations of the benefit of the YMC. \\
~~~~Additionally, the anti-phase oscillation between the flux and affinity also emerges in a model of the glycolytic oscillation \cite{goldbeter1972dissipative}. Since glycolysis is a part of the catabolic process, we cannot adopt the same definition of the thermodynamic efficiency as adopted in the present paper. However, thermodynamic efficiency for the energy conversion, defined similarly, is shown to be improved by the glycolytic oscillation \cite{termonia1981oscillations}. The examples of YMC and glycolytic oscillation may imply the ubiquity of the anti-phase oscillation between the reaction flux and affinity as a mechanism of improvement of the thermodynamic efficiency by enzymatic oscillation.  \\
~~~~In general, intracellular processes include several coupled
reactions that may disequilibrate each other. One potentially effective strategy to achieve higher
efficiency may be to temporally separate these reactions by
switching the abundances of the corresponding enzymes for each reaction in time. This issue will be addressed directly in future extensions of the present model. Furthermore, analyses of several phases in the cell cycle along these same lines may reveal new patterns and mechanisms.

\section*{Acknowledgments}
~~~~The authors would like to thank A. Kamimura, N. Saito, and T. S. Hatakeyama for useful discussions. 
This research is partially supported by the Platform Project for
Supporting in Drug Discovery and Life Science
Research (Platform for Dynamic Approaches to Living System）from Japan
Agency for Medical Research and Development(AMED), and
Grant-in-Aid for Scientific Research (S) (15H05746） from JSPS.

\end{document}